\documentclass[aps,prc,nofootinbib,preprint]{revtex4-1}
\usepackage{graphicx}
\usepackage{amsmath}
\usepackage{amssymb}

\begin{document}
 \title{Missing resonance decays in thermal models}

 \author{V.V. Begun}
 \email{viktor.begun@gmail.com}
 \affiliation{Faculty of Physics, Warsaw University of Technology,
 Koszykowa 75, 00-662 Warsaw, Poland}

 \author{V.Yu. Vovchenko}
  \affiliation{Institut f\"{u}r Theoretische Physik, Goethe Universit\"{a}t Frankfurt,
 D-60438 Frankfurt am Main, Germany }
 \affiliation{Frankfurt Institute for Advanced Studies, Goethe Universit\"{a}t Frankfurt,
 D-60438 Frankfurt am Main, Germany}
 \affiliation{Department of Physics, Taras Shevchenko National University of Kiev,
 03022 Kiev, Ukraine}

 \author{M.I. Gorenstein}
 \affiliation{Frankfurt Institute for Advanced Studies, Goethe Universit\"{a}t Frankfurt, D-60438 Frankfurt am Main, Germany}
 \affiliation{Bogolyubov Institute for Theoretical Physics, 03680 Kiev, Ukraine}

\begin{abstract}
Detailed information on decay channel probabilities is absent for
many high mass resonances, which are typically included in thermal
models. In these cases, the sum over all known decay branching
probabilities is smaller than 1. Due to this systematic
uncertainty of the model, the exact charge conservation may appear
to be violated. We estimate the corresponding number of missing
charge states in the canonical ensemble formulation of the hadron
resonance gas for p+p reactions at the SPS energy $E_{\rm lab}=
158$~GeV: $\Delta B \simeq 0.16$ for baryon charge, $\Delta
Q\simeq 0.12$ for electric charge, and $\Delta S=-0.01$ for
strangeness.
The value of the considered effect is 5-8\%, which seems to be
important enough to include it as a systematic error in the
calculations within a hadron gas.
%
\end{abstract}

\maketitle


The hadron-resonance gas model (HRG) allows to obtain particle
multiplicities at different collision energies with a relatively
good accuracy. In a simplest HRG hadrons and resonances are
assumed to be non-interacting, and full chemical equilibrium is
imposed. This model has just two free parameters - the
temperature, $T$, and the baryon chemical potential, $\mu_B$,
which follow a simple analytic dependence as the function of
collision energy~\cite{Cleymans:2005xv,Vovchenko:2015idt}.
The HRG model works well for nucleus-nucleus (A+A) collisions, and
even for elementary particle reactions $p+p$, $p+\bar{p}$ and
$e^++e^-$~\cite{Becattini:1995if,Becattini:1997rv}
An analytic assumption about the form of the fireball hypersurface
at freeze-out allows further to obtain the $p_T$ spectra of
measured particles for the cost of just one more parameter - the
ratio of the freeze-out radius and
time~\cite{Broniowski:2001we,Begun:2013nga,Begun:2014rsa}.

Many modifications to the standard HRG exist, which improve
agreement with experimental data, in particular, the one at the
LHC. These include sequential
freeze-out~\cite{Chatterjee:2013yga}, a mechanism of proton
suppression due to re-scattering during the
freeze-out~\cite{Becattini:2012xb}, introducing different proper
volumes for different particles~\cite{Alba:2016hwx}, or
considering the possibility of pion
condensation~\cite{Begun:2013nga,Begun:2015ifa}.


%

A list of stable particles and resonances is a key ingredient of a HRG model.
Taking into account more/missing hadron resonances helped with data description at the SPS~\cite{Schnedermann:1993ws}
and the LHC~\cite{Noronha-Hostler:2014aia}.
%
%
The problem is that the list of resonances and their properties is known well only up to
$m\sim 1.5$~GeV.
%
In particular, many decay channels for measured heavy resonances
are unknown. Therefore, the amount of charge that is calculated in
a HRG after including only the known decays of these resonances
listed in Particle Data Tables~\cite{pdg} is different from the
one before the decays~\footnote{Another aspects of missing charges
in p+p reactions were recently reported
in~\cite{Stroebele:2016lkl}.}.
The size of the latter effect can be estimated. For the case of a
proton-proton reaction the charge of the system is known exactly,
$B=Q=2$, $S=0$. If several total $4\pi$ multiplicities are
available experimentally, then their fit in the HRG within the
canonical ensemble gives the thermal parameters: temperature $T$,
system volume $V$, and strangeness undersaturation parameter
$\gamma_S$~(see Ref.~\cite{Vovchenko:2015idt} for details).
%
%
The resulting missing charge for the p+p reactions at laboratory
momentum $158$~GeV/c is $\Delta B/B\simeq 8\%$ for the baryon
charge, $\Delta Q/Q\simeq 6\%$ for the electric charge, and
$\Delta S=-0.01$ for strangeness~\cite{Vovchenko:2015idt}.
%
%
These numbers reflect a strongest effect of the missing decay
channels probabilities for heavy positively charged baryons.

There are alternative ways to deal with the problem of missing
branching ratios. One option is to additionally normalize all the
branching ratios to 100\%~\cite{Torrieri:2004zz,Wheaton:2004qb}.
However, such a normalization produces some error: it artificially
enhances the known channels, and, therefore, suppresses yet
undiscovered channels.
Another option is to assign same/similar branching ratios based on analogies to the nearest states with the
same quantum numbers and known branching ratios~\cite{Andronic:2008gu}.

Therefore, the value of the considered effect is 5-8\%, which
seems to be important enough to include it as a systematic error
in the calculations within a HRG.

\section*{Acknowledgments}
We thank H.~Stroebele for a fruitful discussion. V.Y.V.
acknowledges the support from HGS-HIRe for FAIR. The work of
M.I.G. was supported by the Program of Fundamental Research of the
Department of Physics and Astronomy of National Academy of
Sciences of Ukraine.


\bibliographystyle{h-physrev}
\bibliography{Begun_YSTAR}

%
\end{document}